\documentclass[twocolumn]{aastex7}

\usepackage{amsmath}
\usepackage{mathrsfs}
\usepackage{units} 
\usepackage{threeparttable}
\usepackage{multirow}
\usepackage{soul}
\usepackage{ulem}

\def\eg{{\textit {e.g. }}}

\def\non-RTsim{{\tt non-RTsim}}

\begin{document}

\title{The Importance of the Population III Initial Mass Function in Determining the Characteristics of the Earliest Galaxies}
\shorttitle{The Pop III IMF \& Early Galaxies}
\author{Richard Sarmento}
\email{Richard.Sarmento@asu.edu}
\affiliation{School of Earth and Space Exploration, 
Arizona State University, 
P.O. Box 871404, Tempe, AZ, 85287-1404}

\author{Evan Scannapieco}
\email{Evan.Scannapieco@asu.edu}
\affiliation{School of Earth and Space Exploration, 
Arizona State University, 
P.O. Box 871404, Tempe, AZ, 85287-1404}

\shortauthors{Sarmento and Scannapieco}


\begin{abstract}
We use large-scale cosmological simulations to study the prospect of observing Population III (Pop III) bright galaxies with the {\em James Webb Space Telescope} (JWST). To quantify the impact of radiative transfer (RT), we compare a simulation that includes moment-based RT with one in which RT is handled approximately. Both simulations include a subgrid model of turbulent mixing, which is essential in tracking the formation of Pop III stars.  We find that RT has a minor impact on our results and that  the overall star formation rate densities for both simulations are in fair agreement with observations and other simulations. While our overall galaxy luminosity functions are consistent with current high-redshift observations, we predict a drop of a factor of  at least 6 in detectable galaxy counts at $z = 14$ as compared to $z=12$ at $M_{\rm ab} \le -16$.  Modeling Pop III stars according to a lognormal, top-heavy initial mass function (IMF), we find that these stars contribute no more than  $\approx 1\%$ of the flux of potentially detectable lensed galaxies at $z=12-14$ with $M_{\rm ab} \le -15$. This is because a top-heavy Pop III IMF results in 99\% of Pop III stellar mass being recycled within 10 Myr, well before the $\approx 30$ Myr timescale on which galaxies recover from supernova feedback and heating. These effects conspire to quickly extinguish Pop III star formation, making their detection difficult even for JWST.
\end{abstract}



\section{Introduction}

Despite extensive searches, the first generation of stars remains unobserved. Theoretical studies suggest that such metal-free stars could be detected in the Galaxy if their initial mass function (IMF) extended to masses below $\approx 0.8 M_\sun$ \citep{2006ApJ...653..285S, 2006ApJ...641....1T, 2007ApJ...661...10B,2010MNRAS.401L...5S,2015MNRAS.447.3892H,2016ApJ...826....9I}. However, no one has yet observed such a Population III (Pop III) star in the nearby universe \citep{2002Natur.419..904C, 2004A&A...416.1117C, 2006ApJ...639..897A,2005Natur.434..871F, 2007ApJ...670..774N,2011Natur.477...67C,2014Natur.506..463K,2015Natur.527..484H}. 

At high redshifts, surveys have the potential to observe Pop III stars with higher masses and shorter lifetimes.  Such stars may be detectable through their resultant supernovae \citep{2005ApJ...633.1031S,2011ApJ...734..102K,2012ApJ...755...72H,2020ApJ...894...94R}, through highly-magnified caustic transits of individual star clusters \citep{2018ApJS..234...41W,2022ApJ...940L...1W}, and through measurements of galaxies with significant Pop III stellar populations \citep{2003ApJ...589...35S}.  

The last of these types of observations is the most straightforward, as it can be carried out in the context of high-redshift galaxy surveys \citep{2002ApJ...565L..71M,2004ApJ...617..707D,2006Natur.440..501J,2007MNRAS.379.1589D,2008ApJ...680..100N, 2012ApJ...761...85K, 2013A&A...556A..68C}, and it has already led to the discovery of a $z$ = 6.6 galaxy that displays strong narrow He II $\lambda$1640 emission \citep{2015ApJ...808..139S} -- an indicator of the hard-ultraviolet (UV)  Pop III spectrum  \citep{2001ApJ...550L...1T}. Yet follow-up observations indicate that this system is also explainable by the presence of a narrow-line active galactic nucleus or a young $1/200\, Z_{\odot}$ starburst \citep{2017MNRAS.469..448B,2017MNRAS.468L..77P}, meaning that there has not yet been a confirmed observation of a galaxy whose flux is dominated by Pop III stars.

This may change soon. The \textit{James Webb Space Telescope} (JWST) is greatly expanding our understanding of the high-redshift universe, pushing to higher redshifts and to lower-mass systems that are likely to contain Pop III stars.  However, planning for such observations requires estimating how such galaxies are distributed and, even more importantly, what fraction as a function of magnitude and redshift will be dominated by Pop III flux -- prioritizing them for spectroscopic follow-up. 

For now, we only have general observational clues about the history of  Pop III star formation.  Using extremely deep Hubble Space Telescope (HST) observations, astronomers have been able to produce large galaxy catalogs out to $z=8$ and place initial constraints on galaxy populations out to $z\approx11$ \citep{McLure2013,2013ApJ...762...32C,2013ApJ...773...75O,2015ApJ...803...34B, 2015MNRAS.450.3032M,2016PASA...33...37F, 2016ApJ...816...46M, Rojas-Ruiz2020, Bouwens2021}.  Recently, these measurements have been extended by initial JWST data \citep{Donnan2023,Harikane2023,Furtak2023,Adams2024,Davis2024,Willott2024} which complement existing measurements while pushing out to $z \approx 12.$

At the same time, several groups have used large-scale cosmological simulations and analytic models to investigate galaxy formation, the high-$z$ luminosity function (LF), and galaxy assembly \citep{2012MNRAS.423.1992S, 2015ApJ...807L..12O, 2016ApJ...816...46M,2017MNRAS.469.4863B}. Others have used simulations to explore the transition between Pop III and Population II (Pop II) star formation \citep{2007MNRAS.382..945T, 2007ApJ...654...66O, 2010ApJ...712..435T, 2010MNRAS.407.1003M, 2011ApJ...740...13Z, 2012ApJ...745...50W, 2013ApJ...773..108C, 2013MNRAS.428.1857J,2014MNRAS.440.2498P,Xu2016,Sarmento2018,Jaacks2018,Sarmento2019, 2022ApJ...932...71W,venditti2023needle,Yajima2023}. 

By definition, the first generation of Pop III stars must have formed from primordial gas. However, an initial mass function lacking in low-mass stars may also result from gas with metallicity below a critical threshold, $Z_{\rm crit},$ the exact value of which depends on whether the dominant cooling channel is dust or fine-structure metal lines \citep{2003Natur.422..869S, 2003Natur.425..812B,2005ApJ...626..627O}. While this value is poorly constrained, it is believed to be in the range  $10^{-6} < Z_{\rm crit} < 10^{-3} Z_{\odot}$, which is low enough such that the uncertainty in $Z_{\rm crit}$ has a very weak impact on predictions of the evolution of Pop III stars \citep{Sarmento2019}. 

Instead, the transition from Pop III to Pop II star formation is governed by the transport and mixing of SN ejecta into primordial gas \citep{Sarmento2016}, a process that depends on the combined effects of supernovae and turbulence.  Our previous work has been focused on tracking this transition, developing an approach that allows us to track the effects of subgrid turbulent mixing in each resolution element \citep{2012JFM...700..459P,2013ApJ...775..111P}, estimating the fraction of Pop III stars created in regions that would otherwise be considered polluted above $Z_{\rm crit}$.  This can lead to an increase of at least $\approx 3$ times the Pop III star-formation rate density as compared to similar cosmological simulations that do not account for subgrid mixing \citep{Sarmento2016}. 

At the high redshifts currently being probed by JWST, this picture is further complicated by the presence of radiative feedback, due to ultraviolet radiation in both the H-ionizing and Lyman-Werner (LW) bands.  Hydrogen-ionizing photons, in particular, provide an important heating source that inhibits the formation of stars in low-mass galaxies \citep{Dawoodbhoy2018,Katz2020}. This suppression has been studied in numerous dedicated radiative-transfer simulations that simultaneously track the formation of galaxies and the resulting feedback \citep[e.g.][]{Iliev2007,Gnedin2014,Ocvirk2015,Pawlik2017,Ocvirk2020}.

Recent advances by our group have, for the first time, allowed these self-consistent radiative transfer (RT) simulations to be combined with the subgrid turbulent mixing model needed to accurately trace Pop III star formation \citep{2022ApJ...935..174S}.  Our approach uses a customized version of the \textsc{Ramses-RT} \citep{2010ascl.soft11007T} cosmological adaptive mesh refinement (AMR) code to follow reionzation, galaxy formation, and enrichment from the dawn of star formation at $z\approx 20$, to $z\approx7$. In \cite{2022ApJ...935..174S}, we focused our analysis on a small simulation volume, 3 $h^{-1}$ comoving Mpc (cMpc) on a side, comparing it with a non-RT simulation to understand the physical differences that arise when reionization and the Pop II/Pop III transition are modeled simultaneously.

Here we describe the results of a new, larger-scale simulation 24 $h^{-1}$  cMpc on a side, which allows us to simulate galaxies at the luminosities and redshifts being targeted by current high-redshift galaxy surveys.  In particular, we generate rest-frame UV ($1500$ \AA) galaxy luminosity functions at $z=8$ to $14,$  compare them with existing photometric data, and use them to predict upcoming data sets.  Combining these results with our unique capability to track the rate of subgrid metal pollution, we are then able to make predictions as to the galaxy luminosities and redshifts that are most likely to show Pop III features, such as narrow He II $\lambda$1640 emission, when they are followed up spectroscopically.

The work is structured as follows.  In Section 2, we describe our methods, including a brief discussion of the implementation of our subgrid model for following the evolution of the pristine gas fraction, our approach to halo finding, and the spectral energy distribution (SED) models used to compute the luminosity of our stars. In Section 3, we show that our high-redshift LF agrees with current observations and we make predictions for future JWST surveys. We also compute the fraction of Pop III flux emitted by these early galaxies that can be used to guide the search for metal-free stars. Conclusions are given in Section 4.

\section{Methods}

We used a customized version of \textsc{Ramses-RT} \citep{2002A&A...385..337T, Rosdahl2013, Rosdahl2015a}, a cosmological adaptive mesh refinement (AMR) code that includes coupled radiation hydrodynamics (RHD), to generate the results presented here. Our version of \textsc{Ramses-RT} tracks the unpolluted fraction of gas in each simulation cell using a subgrid estimate of turbulence. This allows us to account for Population III star formation in otherwise polluted cells \citep{Sarmento2016}. For this work we ran two simulations, one with RT and one without (henceforth \textit{non-RT}), from $z=150$ to $z\approx 7$.  In this section, we describe the key features of the code, along with our simulations' parameters and analysis methods.

\subsection{RT Simulation}

\textsc{Ramses-RT} models the evolution of dark matter, baryons, and stars via gravity and hydrodynamics, along with radiative transfer and non-equilibrium radiative heating and cooling. Our RT simulation tracks the ionization fractions of hydrogen, helium, and molecular hydrogen \citep{Nickerson2018} in each cell. We grouped ionizing photon energies into the following bins: Lyman Werner H$_2$ dissociating, H ionizing, and 2 levels for He ionizing radiation:
\begin{align*}
 11.20\,{\rm eV} &\leq \epsilon_{\rm LW} < 13.60\,{\rm eV}, \\
 13.60\,{\rm eV} &\leq \epsilon_{\rm HII} < 24.59\,{\rm eV},\\
 24.59\,{\rm eV} &\leq \epsilon_{\rm HeII} < 54.42\,{\rm eV}, \,\, {\rm and} \\
 54.42\,{\rm eV} &\leq \epsilon_{\rm HeIII}.
\end{align*}  
For the RT simulation, we modeled the ionization fractions $x_{\rm HII}$, $x_{\rm HeII}$, $x_{\rm HeIII}$ for each cell self-consistently with the temperature and radiation field in terms of photon density and the flux for each photon group. 

\textsc{Ramses-RT} uses moment-based\footnote{For information and comparisons on the various methods used to model radiative-transfer see \cite{2009MNRAS.400.1283I}.} radiative-transfer to compute photon densities and fluxes {\citep{Rosdahl2015a}.  The resulting ionization states are advected between cells. Our simulation used the {\it on-the-spot approximation} \citep{Rosdahl2013} such that recombination photons were reabsorbed in the same cell in which they were emitted. Time-step management was handled by employing a reduced speed of light approximation \citep{Ocvirk_2019}. We adopted 0.01$c$ for this work. Star particles (SPs) are the sole source of photons and they are based on externally-specified SED models. We used {\it Starburst99} \citep{2011ascl.soft04003L} for star particles with $4\times10^{-4} \le Z \le 0.05,$ and models by \cite{Raiter2010} for star particles with metallicities $Z < 4\times10^{-4}.$ Our SEDs modeled SP ages between 10 kyr and a Gyr. 
The RT and non-RT simulations share the settings described below. 

\subsection{Scalar advection}

We used the Harten--Lax--van Leer contact (HLLC) Riemann solver \citep{Toro:1994we} to advect the typical cell-centered gas variables, the RT ionization states for the RT simulation, and the scalars added by our team \citep{Sarmento2016}. These scalars track the turbulent velocity, $v_t$, the pristine gas mass fraction, \textbf{$P$}, and the metals generated by Pop III supernovae (SNe), $Z_P$, as distinct from Pop II SN metals, $Z$. While not utilized in this study, associating two different metallicity scalars with the ejecta from Pop III and Pop II stars allows our team to explore the effect of different metal abundances for each type of SN on the observational characteristics of subsequent stellar generations. This can all be done in post-processing.

The pristine gas fraction tracks the unmixed fraction of gas in each simulation cell as
\begin{equation}\label{eqn:P}
\frac{dP}{dt} = -\frac{n}{\tau_{\rm con}}P(1-P^{1/n}),
\end{equation}
where $n$ is based on the turbulent Mach number of the gas and $\tau_{\rm con}$ is the convolution time scale, which is based on the cell size and the gas' turbulent velocity. The underlying subgrid model was developed by \cite{2013ApJ...775..111P}, see also \cite{2010ApJ...721.1765P, 2012JFM...700..459P}. Tracking this scalar improves the estimate of the cosmic Pop III star formation rate by reducing the impact of numerical diffusion.

\subsubsection{Fix to the Turbulent Velocity}\label{sec:fix}

During this study we discovered an error in the computation of the turbulent velocity, $v_{t}$, that affects previous studies that used our modified version of \textsc{Ramses}. An incorrect value was used in the computation that resulted in a significant overestimation of $v_{t}.$ This resulted in an {\it under-estimation} of the mixing time since the turbulent velocity, $v_t$, is inversely proportional to the convolution time, $\tau_{\rm con}$,
\begin{equation}
\tau_{\rm con} = \frac{\Delta x}{v_{\rm t}},
\end{equation}
where $\Delta x$ is the simulation cell size.
Since $dP$ is inversely related to $\tau_{\rm con}$ the change in the pristine fraction of the gas was previously {\it overestimated,} resulting in an overly rapid decay of the pristine gas in each simulation cell. 

The impact of this error is that our previous works {\em underestimated} the number of Pop III stars that could form in incompletely mixed gas.  Note that standard simulations without a subgrid model assume instantaneous mixing of metals into the volume of each simulation cell.  Hence the results in \cite{Sarmento2016, Sarmento2018, Sarmento2019, 2022ApJ...935..174S} related to the Pop III star formation in early galaxies should be considered a lower-bound, which accounts for some, but not all of the delay in the Pop III to Pop II transition due to mixing at subgrid scales.
\subsection{Star Formation}

\textsc{Ramses-RT} creates collisionless SPs according to a \cite{Schmidt59} law when the local density, $\rho_{\rm gas}$, exceeds the density threshold, $\rho_{\rm th}=  1.12\, H\, {\rm cm}^{-3}$, and is also at least 200 times the average gas density, $\overline{\rho}_{\rm gas}$, in the simulation 
\begin{equation}\label{eqn:sf}
\dot{\rho_{\star}} =  \epsilon_{\star} \frac{\rho_{\rm gas}}{t_{\rm ff}} \theta(\rho_{\rm gas}- \rho_{\rm th}) \theta(\rho_{\rm gas}- 200\overline{\rho}_{\rm gas}).
\end{equation}
Here, $\epsilon_{\star}=0.03$ is the star formation efficiency, $t_{\rm ff}$ is the gas free-fall time, and the Heaviside step function, $\theta$, ensures star formation only occurs in collapsed objects when the density exceeds our threshold \citep{Rasera2006, Trebitsch2017}.

Each SP represents an initial mass function (IMF) of stars and evolves via a particle-mesh solver with cloud-in-cell interpolation \citep{Guillet_2011}. The SP mass was set by the star-forming density threshold and our resolution resulting in $m_{\star} = {4.3 \times 10^{4}\,} h^{-1}\, M_{\odot}$. The final mass of each SP was drawn from a Poisson process such that it is a multiple of $m_{\star}$. 

Given our SP mass and the approach to clump finding outlined below we were able to reliably identify galaxies down to a stellar mass of approximately $5\times 10^6 M_\odot$. However, the algorithm does identify stellar clumps below this threshold and we included them in the analysis even though stellar clusters with masses below about $10^5 M_\odot$ do not contribute to the observable part of the LF.

We assumed an ideal gas with a ratio of specific heats $\gamma$ = 5/3. The critical metallicity, marking the distinction between Pop III and Pop II star formation was set to $Z_{\rm crit} = 10^{-5} Z_{\odot}$. We have explored the sensitivity of Pop III star formation to this value and found it to be largely insensitive over the range $10^{-4} \le  Z_{\rm crit}/ Z_{\odot} \le 10^{-6}$ \citep{Sarmento2019}.

\subsection{Feedback}\label{sec:feedback}

We implemented a Salpeter IMF for Pop II stars with $Z > Z_{\rm crit}$ in both simulations. Ten percent (10\%) of each Pop II SP's mass represents stars more massive than 8 $M_\odot$ that go  SN after 10 Myr \citep[e.g.][]{Raskin_2008, Somerville2008}. Pop III stars ($Z \le Z_{\rm crit}$) were modeled using a top-heavy log-normal distribution with masses ($1 \le M_\star/M_\odot \le 500$), such that 99\%, by mass, explode within 10 Myr \citep{1973MNRAS.161..133L, tumlinson2006chemical, Raiter2010} of formation. This Pop III IMF resulted in a very high fraction of stellar mass being recycled into the interstellar medium (ISM) shortly after these stars were formed. 

We used a SN energy $E_{\rm SN}=\bf{{5}}\times 10^{51}$ ergs for each 10 $M_{\odot}$ of exploding stars, for all stars formed throughout the simulation.  Hence, a Pop II SP with a mass of 1000 $M_\odot$, using a Salpeter IMF, will generate $10\times(5\times 10^{51})$ ergs 10 Myr after the SP is born. While slightly higher than the normal energy used for SN, our choice helps to model the effect of SN in the pair-instability range where energies are typically closer to $10^{53}$ erg: i.e. -- Stars in the mass range $100\rightarrow300 M_\odot$ generate $E_{\rm SN} = 4\times 10^{52} \rightarrow1.2\times 10^{53} $ erg. The added feedback also helps to keep the overall SFRD in rough agreement with high redshift observations and is used to mitigate the overcooling problem \citep[e.g.][]{Rosdahl2017,Rosdahl2018}. Finally, the mass fraction of metals generated by all SNe is set to 15\%  even though metal yields may have been higher for Pop III stars \citep{2003ApJ...589...35S,2005ApJ...624L...1S}.

Our choice of Pop III IMF was motivated by \cite{Sarmento2019}. In that work, we found a good match between the nucleosynthetic products of a top-heavy log-normal IMF and the metals detected in Milky Way (MW) CEMP-no stars \citep{Yoon2016}. As such, our choice of Pop III IMF has a large effect on the surviving fraction of Pop III stars  as well as the metallicity of early galaxies. Other recent authors argue for a smaller mass range \citep{2020ApJ...892L..14S, 2022ApJ...937L...6R}, but there remains a large amount of uncertainty concerning both the IMF and SN metal yields of the first stars. 

For the RT simulation, we set the photon escape fraction, $f_{\star, {\rm esc}} = 2.0$. This parameter should not be confused with the physical, \textit{galactic} escape fraction. The parameter $f_{\star, {\rm esc}}$ is a {\sc Ramses} input value that scales the number of photons generated by each SP and that escape the star-forming cell.  As such, it is a free parameter used to tune the SED feedback \citep{Rosdahl2018, 10.1093/mnras/stw2869}.  In particular, it helps model the effects of unresolved structures such as chimneys in the ISM allowing photons to escape at a higher rate than the local average gas density would typically allow.  This parameter is also used to help overcome the overcooling problem in cosmological simulations.  Note that it has no effect on the post-simulation generation and analysis of galaxy luminosities. See \S{\ref{GSM}} for details on the luminosity model used in this work.

Finally, we did not model black hole growth in our simulations since their feedback is not likely significant for early galaxy formation at faint and moderate luminosities \citep{Scannapieco_2004,Somerville2008}.  However, this assumption is coming under new scrutiny given recent results \citep[e.g.][]{2023ApJ...957L...7K}.

\subsection{Simulation Parameters}

As in our previous work, we adopted cosmological parameters from \cite{Komatsu_2011} with $\Omega_{\rm m} = 0.267$, $\Omega_{\Lambda} = 0.733$, $\Omega_{\rm b} = 0.0449$, $h = 0.71$, $\sigma_8 = 0.801$, and $n = 0.96$.  The parameters $\Omega_{\rm m}$, $\Omega_{\Lambda}$, and $\Omega_{\rm b}$ are the total matter, vacuum, and baryonic densities, in units of the critical density,  $h$ is the Hubble constant in units of 100 km/s; $\sigma_8$ is the variance of linear fluctuations on the 8 $h^{-1}$ Mpc scale; and $n$ is the ``tilt" of the primordial power spectrum \citep{Larson_2011}. 

For both simulations, we modeled early galaxy formation using a 24 $h^{-1}$ cMpc on-a-side cube, evolved down to $z\approx 7$. The patchy nature of reionization together with cosmic variance means that our simulations may not capture samples of rare, bright galaxies at $z\ge10$ \citep{Furlanetto2004}. Reliably probing the simulated LF down to densities $\phi < 10^{-5} \,{\rm mag}^{-1}\,{\rm cMpc}^{-3}$ requires simulations at least 16$\times$ larger \citep[e.g.][]{Iliev2014}.  Conversely,  our simulations are well suited to probe the faint end of the LF down to magnitudes accessible by JWST.

The initial grid scale and simulation size set the dark matter (DM) particle mass. For this work, $M_{\rm DM} = 1.1 \times 10^{6}\, M_{\odot}$. We set the initial refinement level to $\ell_{\rm min} = 10,$ corresponding to a coarse, initial grid resolution $\rm \Delta x_{\rm max} = 23\; h^{-1}$ comoving kpc (ckpc) in both simulations. This is a compromise that provides reasonable resolution in the intergalactic medium without creating an excessive computational load. 

We used a quasi-Lagrangian approach to refinement such that cells were refined as they became approximately 8$\times$ overdense. This strategy attempts to keep the amount of mass in each cell roughly constant as the simulation advances. We set the maximum refinement level to $\ell_{\rm max} = 18,$ resulting in a best average resolution of 91.6 $\rm h^{-1}$ comoving pc (cpc) in the densest, most refined regions.

Each simulation started with the same initial conditions. They were generated using the Multi-Scale Initial Conditions code \citep[MUSIC, ][]{2013ascl.soft11011H}. The initial gas metallicity was $Z = 0$ and we define $Z_{\rm crit} = 10^{-5} Z_\odot$, the boundary between Pop III and Population II (Pop II) star formation. 

The nonlinear length scale at the end of the simulation, $z= 6.7$, was 25 h$^{-1}$ ckpc , corresponding to a mass of $4.9\times 10^{6}$ h$^{-1}$ $M_\odot$.  For the non-RT simulation we set the epoch of reionization to $z=8.5$.

\subsection{Clump Finding and Primordial Galaxies}

We used the RAMSES Parallel HiErarchical Watershed (PHEW) clump finder \citep{2015ComAC...2....5B} to find galaxies based on SP densities. We used star particles instead of DM particles since we are mainly interested in luminous objects and not the characteristics of extended DM halos. The algorithm was used in both runs and executed concurrently with the evolution of the simulations. Table \ref{tab:PHEW} identifies the parameters used by PHEW.

\begin{deluxetable}{r|l}
\tablewidth{150px}
\tabletypesize{\footnotesize}
\tablecolumns{2} 
\tablecaption{\label{tab:PHEW}PHEW parameters. } 
\tablehead{\colhead{~~~~~~~~~~~~~~Parameter Name} &  \colhead{Value~~~~~~~~~~~} } 
\startdata
relevance\_threshold & 3\\ \hline 
density\_threshold & 80  \\ \hline  
saddle\_threshold & 100  \\ \hline  
mass\_threshold & 100 \\ \hline  
ivar\_clump & 0  \\
\enddata 
\end{deluxetable} 

Briefly, the algorithm finds SP density peaks and saddle points between them. The $density\_threshold$ specifies the required minimum overdensity for a clump to be considered by the algorithm. The $saddle\_threshold$ defines the density below which two neighboring clumps can be considered independent. If a saddle point between two peaks is above this user-specified threshold, it is merged into the more dense neighbor. The $relevance\_threshold$ is the ratio of the peak density to the saddle density. Only clumps with a peak-to-saddle ratio above $relevance\_threshold$ are considered above background noise. Similarly, only clumps with $mass\_threshold \times m_{\star}$, $\approx$100 SPs, are saved and written to disk.  This results in galaxies with $\gtrsim 10^6 M_\odot$ in stars. Note that due to variability in SP mass some clumps have less than $100\times m_{\star}$.

Setting $ivar\_clump = 0$ tells the algorithm to use  particle density, as opposed to the gas density, for clump finding.  We consider all SP in the clump finder, regardless of their age. This means we include stellar remnants in the analysis that, at least for Pop III stars, contribute very little mass to the clump. For a complete description see \cite{2015ComAC...2....5B}.

 The earliest Pop III stars are thought to form in halos below about $10^6 M_\odot$, the so-called minihalos, that rely on molecular hydrogen for cooling. Subsequent, larger galaxies may be enriched via mergers with these objects raising their average metallicity above the Pop III metallicity cutoff.  However, as will be discussed later, our choice of IMF for Pop III stars contributes metals, more quickly, to the gas than is assumed by a Salpeter or Kroupa IMF. Furthermore, the contribution of early minihalos to the metallicity of larger structures is greatly reduced as they depend on H2 molecules to cool and the primordial molecular hydrogen in our simulation volume is quickly destroyed soon after the first stars are born \citep{Sarmento2016}.

\subsection{Galaxy Spectral Models}\label{GSM}

The rest-frame UV luminosities of our SPs, for each simulation,  were based on a set of simple stellar population SED models parameterized by the SPs' ages, metallicities, and masses.  The SEDs were based on \textit{STARBURST 99} \citep{ 2014ApJS..212...14L}, henceforth \textit{SB99}, for Pop II SPs.  Pop III SPs used \cite{Raiter2010} and \cite{2003A&A...397..527S}, henceforth \textit{R10}, SEDs.  We assumed Pop II stars with $Z_{\star} > Z_{\rm crit}$ follow a \cite{1955ApJ...121..161S} IMF  with masses between 0.8 and 100 $M_\odot$.  We modeled Pop III SPs   with $Z_{\star} \leq Z_{\rm crit}$ based on the \textit{R10} SEDs for a zero-metallicity population,  with a log-normal IMF  centered on a characteristic mass of $60M_{\odot}$ with $\sigma=1.0$ and a mass range $1\, M_{\odot} \le M_\star \le 500\, M_{\odot}.$  As mentioned above, this Pop III IMF provided the best match between the observed abundances of (Pop II) MW CEMP-no halo stars and the IMFs explored in \cite{Sarmento2019}. The SEDs model instantaneous bursts across the age range of SPs in the simulation for both types of stars.

Our subgrid model gives us the ability to estimate the metal-free mass fraction of SP formed in cells that are incompletely mixed. Hence, the Pop III stellar mass includes not only stars formed in cells that are metal-free, but those with $P_{\star} > 0$. Specifically $P_{\star}\times M_{\star}$ represents the Pop III mass fraction of each SP.  Hence the total mass of Pop III stars in our galaxies is

\begin{equation}\label{eq:pop3mass}
\begin{aligned}
M_{\star, \rm III} = \sum_{n=1}^{N}{P_{\star, n} \; M_{\star, n}} ,
\end{aligned}
\end{equation}
where $N$ is the total number of SPs in a galaxy and $M_{\star, n}$ is the mass of each SP. Note that $0\le P_{\star, n} \le 1$ for each SP in the simulation.

To compute the observational flux, we redshifted each of our SEDs over the range $z$=6 to 16, applying Lyman forest and continuum absorption as described in~\cite{1995ApJ...441...18M}.  This transforms the rest frame \textit{SB99} and \textit{R10} SEDs (erg/s/\AA$/M_{\odot}$) into observational fluxes (erg/s/Hz/cm$^2/M_{\odot}$) across the parameter space of our SPs:
\begin{equation}\label{eq:LtoF}
\begin{aligned}
f(\nu,z) =  \frac{L_\nu(\nu_e)}{4 \pi D_L^2} (1+z) \, \mathcal{M}(\nu_{o}, z),
\end{aligned}
\end{equation}
where $\nu_{o}$ and $\nu_{e}$ are in Hz for the observed and emitted reference frames, respectively; $D_L$ is the luminosity distance; and $\mathcal{M}(\nu_{o}, z)$ is the~\cite{1995ApJ...441...18M} Lyman absorption function. We also used this formulation to compute the absolute magnitudes by setting $z=0$, $D_L = 10$ pc and $\mathcal{M}(\nu_{o}, z)=1.0$ in equation (\ref{eq:LtoF}).

\begin{figure*}[t]
\begin{center}
\includegraphics[width=2.0\columnwidth]{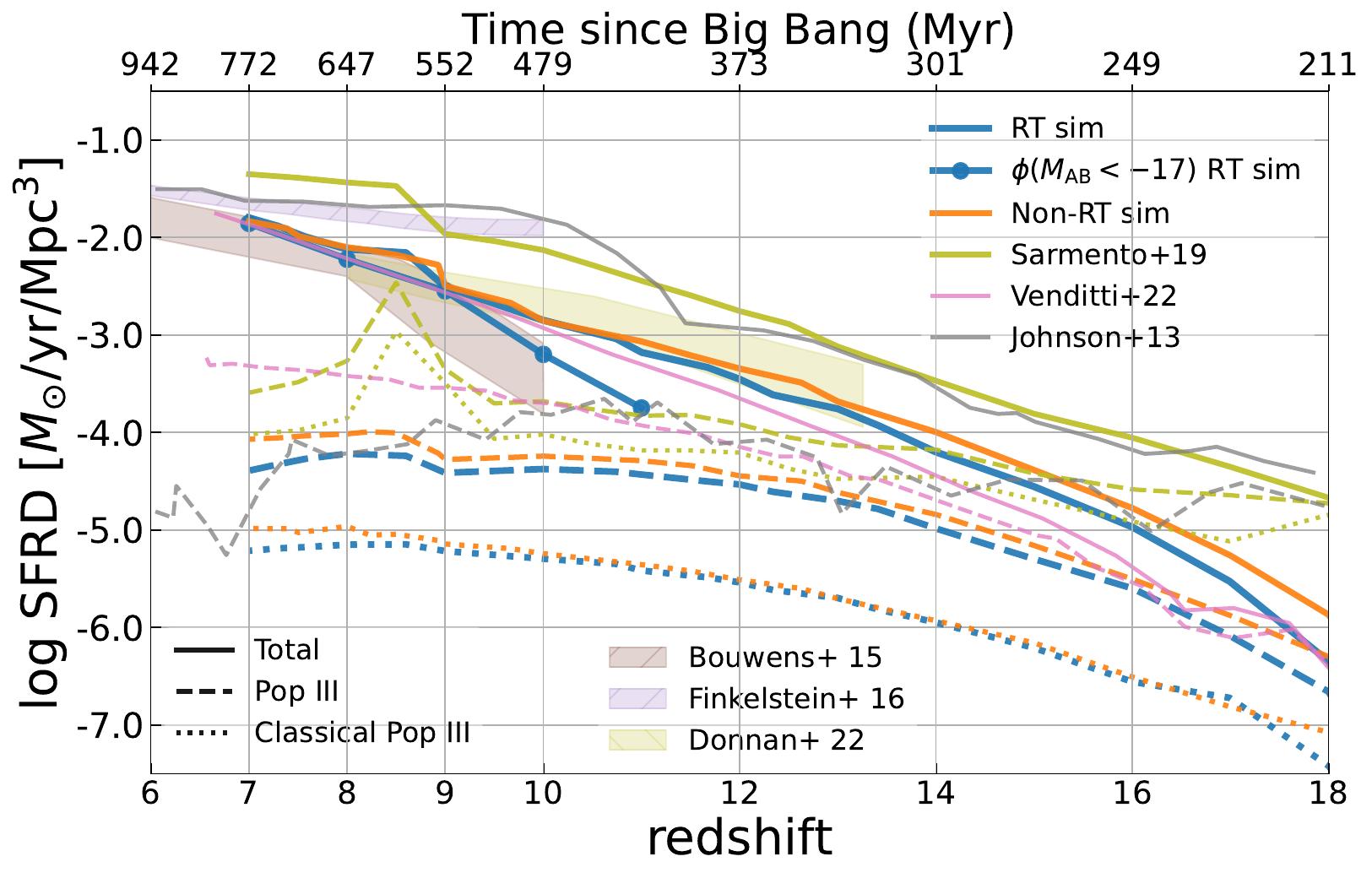}
\caption{The comoving SFRD for the RT (blue lines) and non-RT (orange lines) simulations. We include observations by \cite{2015ApJ...803...34B, Finkelstein_2016, Donnan2023} (shaded regions). The  blue line with large dots is a LF-derived SFRD for $M_{\rm AB} < -17$. Our LF-derived SFRD is in fair agreement with recent observations. We also include SFRDs for particle-based simulations by \cite{venditti2023needle, 2013MNRAS.428.1857J} for comparison. The olive lines are from \cite{Sarmento2019}: a simulation with the same resolution but lower feedback as compared to these simulations. The light dotted lines indicate the Classical Pop III SFRD. This is the Pop III rate that does not include our subgrid mixing model. It results in less than 10\% of the Pop III rate as compared to our subgrid model. See the text for discussion.}
\label{fig:sfrd}
\end{center}
\end{figure*}

\begin{figure}[t]
\begin{center}
\includegraphics[width=1.0\columnwidth]{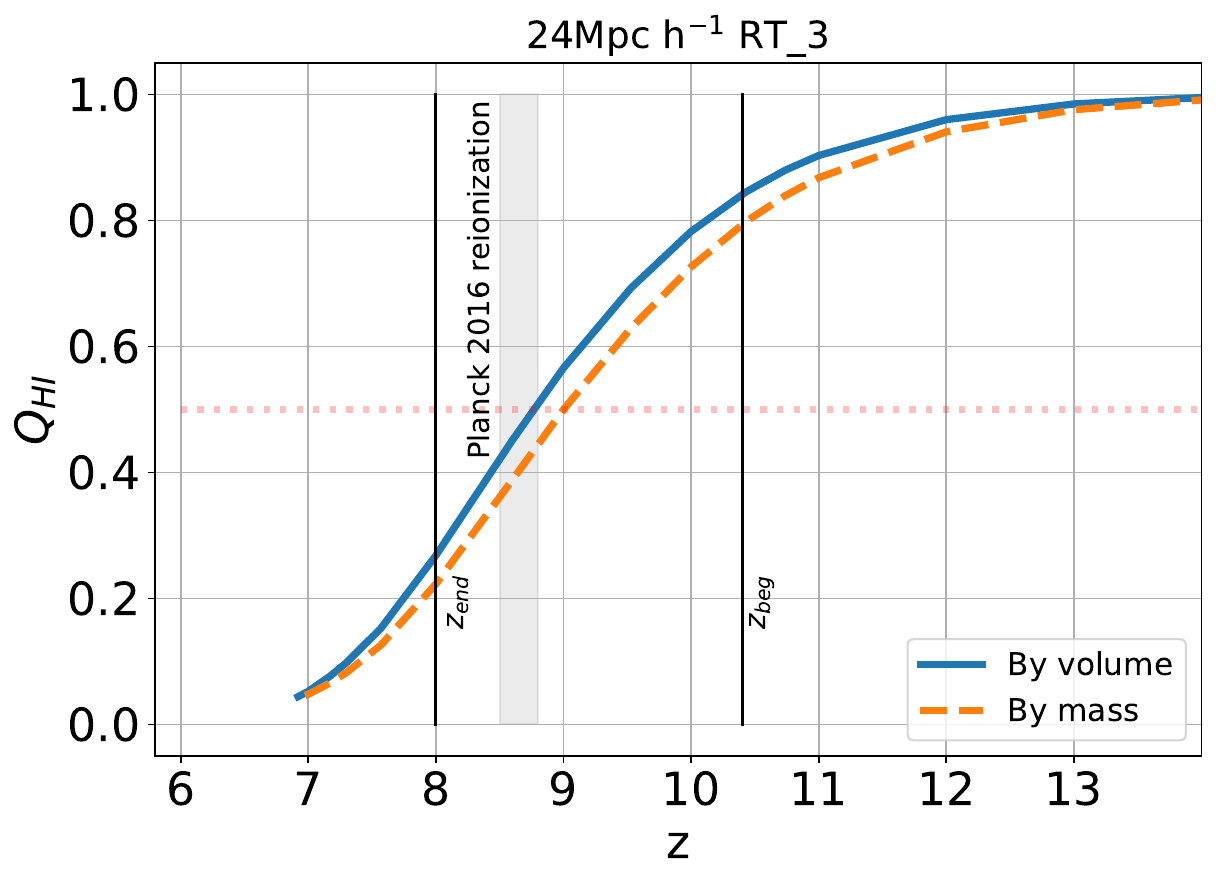}
\caption{The ionization history of our RT simulation, as a function of volume and mass, with limits from \cite{2016A&A...596A.108P} using the prior ``$z_{end} > 6$'' from that work. The red dotted line indicates 50\% ionization, which indicates that the simulation is 50\% ionized at $z \approx 9$. This result is consistent with Planck constraints.  Here we depict the beginning and ending of reionization for the aforementioned prior. The non-RT simulation turns on a UV background at $z=8.5$ to simulate the effects of reionization.}
\label{fig:Q_HI}
\end{center}
\end{figure}

\section{Results}
\subsection{Star Formation \& Reionization}

Next, we  compare the characteristics of our simulated galaxies over the range $8 \le z \le 14$. We focus on the detectability of the galaxies by JWST and the differences between the RT and non-RT simulations. Figure \ref{fig:sfrd} depicts the star formation rate density (SFRD) for our simulations, along with observationally-derived SFRDs from \cite{2015ApJ...803...34B}, \cite{Finkelstein_2016}, and \cite{Donnan2023}. We also include SFRDs from (particle-based) simulations by \cite{venditti2023needle} and \cite{2013MNRAS.428.1857J}. 

Our overall SFRDs, which include all the star particles formed in the simulation, agree with observations. Note that observational SFRDs are based on galaxies that are typically more luminous than $M_{AB} \approx -17$ \citep{2015ApJ...808..104O}. Counting simulation stars only above this threshold results in our ``Luminosity Function-derived'' SFRD (blue line with large dots) that is consistent with \cite{2015ApJ...803...34B}.

Our overall SFRDs are  close to \cite{venditti2023needle} at $z\leq 12$ but larger at higher redshifts.
This is explained by the lower resolution of the \cite{venditti2023needle}, which  has a DM particle mass of $3.53 \times 10^{7}\, h^{-1}\, M_\odot$ or approximately $45$ times more massive than the DM particles in our simulation. This lower resolution results in a delayed start to star formation as compared to our simulation. In turn, this delay leads to an increase in the Pop III star formation rate density in that simulation, which accounts for 50-60\% of stars formed through $z=14$ and which continues to outpace our Pop III SFRDs at lower redshifts. Lastly, \cite{venditti2023needle} does not model stellar radiative transfer  enhancing that simulations Pop III SFR as compared to ours at $z<12$.

We also include  SFRDs from \cite{2013MNRAS.428.1857J}, another particle-based RT simulation with a much smaller volume (4 Mpc $h^{-1}$ on a side) and higher resolution. That simulation has a DM particle mass of $6.16 \times 10^{3}\, M_\odot$ ($\approx$8 ckpc). Our overall SFRD  is lower than} the star formation rate in that work largely due to its ability to trace star formation in smaller galaxies. Our increased SN and stellar feedback also likely play a role here. Our Pop III SFRD is also lower than theirs at high redshift, but comparable from $9 \le z \le 14.$  Note that \cite{2013MNRAS.428.1857J} includes the radiative background contribution of distant sources, while our work only includes the radiation generated from stars within the simulation volume. Hence, at later times ($z\lesssim 9$), radiative feedback in \cite{2013MNRAS.428.1857J} more effectively suppresses star formation in smaller halos \citep{2008ApJ...673...14O}  resulting in a Pop III SFRD that falls off faster than ours. 

The olive lines are from \cite{Sarmento2019}, a simulation with the same resolution, no radiative feedback, and 1/8 the volume.  The most important difference between this work and the current one is the difference in stellar feedback: the current simulations uses 2x the SN energy per unit mass and include the effects of radiative transfer. The added feedback, in both scale and type, accounts for the differences in the SFRDs.
 
 Finally, in Fig.\ \ref{fig:sfrd} we also include curves (dotted lines) for our simulations' Pop III SFRDs that do not include the effects of the subgrid mixing model. We call this the ``Classical Pop III SFRD''. Our data indicate that more than 10x the mass in Pop III stars is generated in cells where mixing is incomplete as compared to unpolluted gas cells.

\begin{figure*}[t]
\begin{center}
\includegraphics[width=\textwidth]{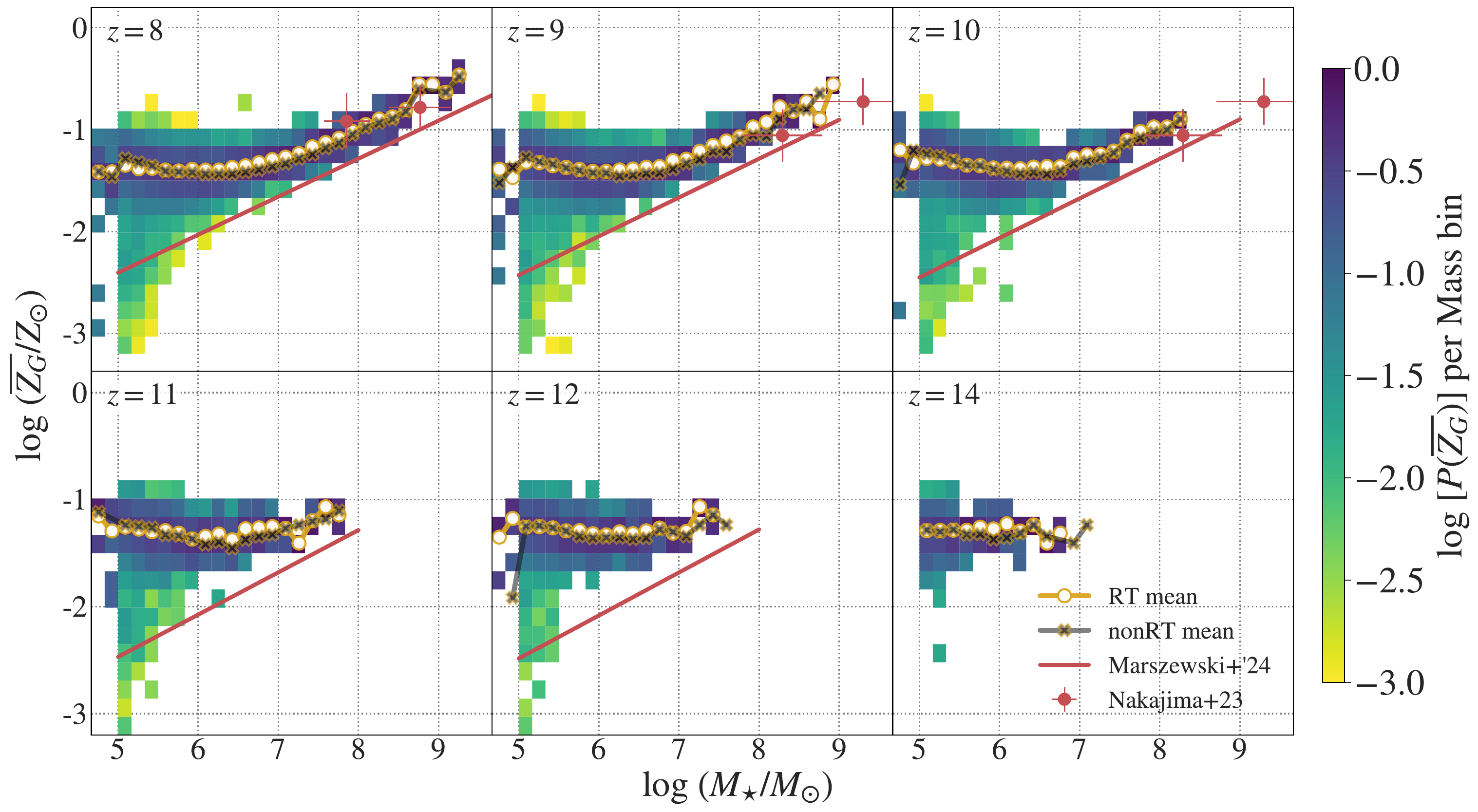}
\caption{The normalized probability, in each mass bin, of finding a galaxy with a metallicity in the range $1 < \big<Z_{\rm G}\big>/Z_{\odot} \le 5\times 10^{-4}$ for our RT simulation. The tan line with circles identifies the mean metallicity across the mass range indicated. The grey line with x's is the mean for the non-RT simulation. The red line is from \cite{2024ApJ...967L..41M} and the FIRE simulation. The red dots, with error bars, are from \cite{2023ApJS..269...33N} and based on JWST observations. See the text for a discussion. }
\label{fig:gMZ}
\end{center}
\end{figure*}

Our RT and non-RT simulations' SFRDs are very similar, although the non-RT simulation shows a slight increase in both the overall and Pop III SFRDs. Specifically, the overall non-RT SFRD is about 20\% above the RT simulation. This increases to 25-35\% for the Pop III rate. This demonstrates the modest effect radiative feedback has on  star formation and early galaxy development. 

Figure~\ref{fig:Q_HI}, depicts the reionization history of our RT simulation along with limits from \cite{2016A&A...596A.108P}. The results are in good agreement with current estimates \citep[e.g.][]{Jung_2020}, although the patchy nature of reionization often makes it difficult to precisely replicate the reionization timeline in moderately sized volumes.

\subsection{The Galaxy Mass-Metallicity Relation}

 It is thought that most Pop III stars in the very high redshift universe form in molecular cooling halos with masses below $\approx 5\times 10^{6} M_\odot$, due to the presence of trace levels of H$_2$ and HD after recombination. Later galaxies may be polluted by mergers with these objects resulting in pre-enrichment, lowering the Pop III SFR. Our simulation resolves galaxies with masses down to about $10^6\,M_\odot$. We partially compensate for the effects of mergers with unresolved galaxies via our use of a top-heavy IMF. Our choice of IMF for Pop III stars contributes $\approx$ 10 times more metals to the gas than is typically assumed by a Salpeter or Kroupa IMF. This rapid pollution of galaxies raises the average metallicity of our galaxies across the mass range resolved.

Figure \ref{fig:gMZ} displays the mass-metallicity relation (MZR) for our simulations' galaxies. It depicts the normalized probability, per mass bin, of finding galaxies across the metallicity range shown for the RT simulation. The dotted and crossed lines depict the means for the RT and non-RT simulations.

Both our RT and non-RT show little evolution in the mean MZR as a function of mass at $z>10$.  At $z<10$ there is only a moderate increase in metallicity at the top end of galaxy masses. This is in contrast to simulations by \cite{Sarmento2018} and \cite{2024ApJ...967L..41M}. Both of these works display a positive correlation between metallicity and galaxy stellar mass. The primary cause of this difference is, once again, the choice of Pop III IMF. 

The simulation in \cite{Sarmento2018} used a Salpeter IMF for all stars in the simulation. This results in 10\% of the mass fraction of new SPs going SN after 10 Myr. Similarly the FIRE simulations by \cite{2024ApJ...967L..41M} used the Kroupa IMF \citep{2001MNRAS.322..231K}. These choices drastically reduced the amount of metals injected by SN as compared to this work. In contrast, the Pop III SN rate for both the RT and non-RT simulations was 99\% after 10 Myr. 

\begin{figure*}[!t]
\begin{center}
\includegraphics[width=2\columnwidth]{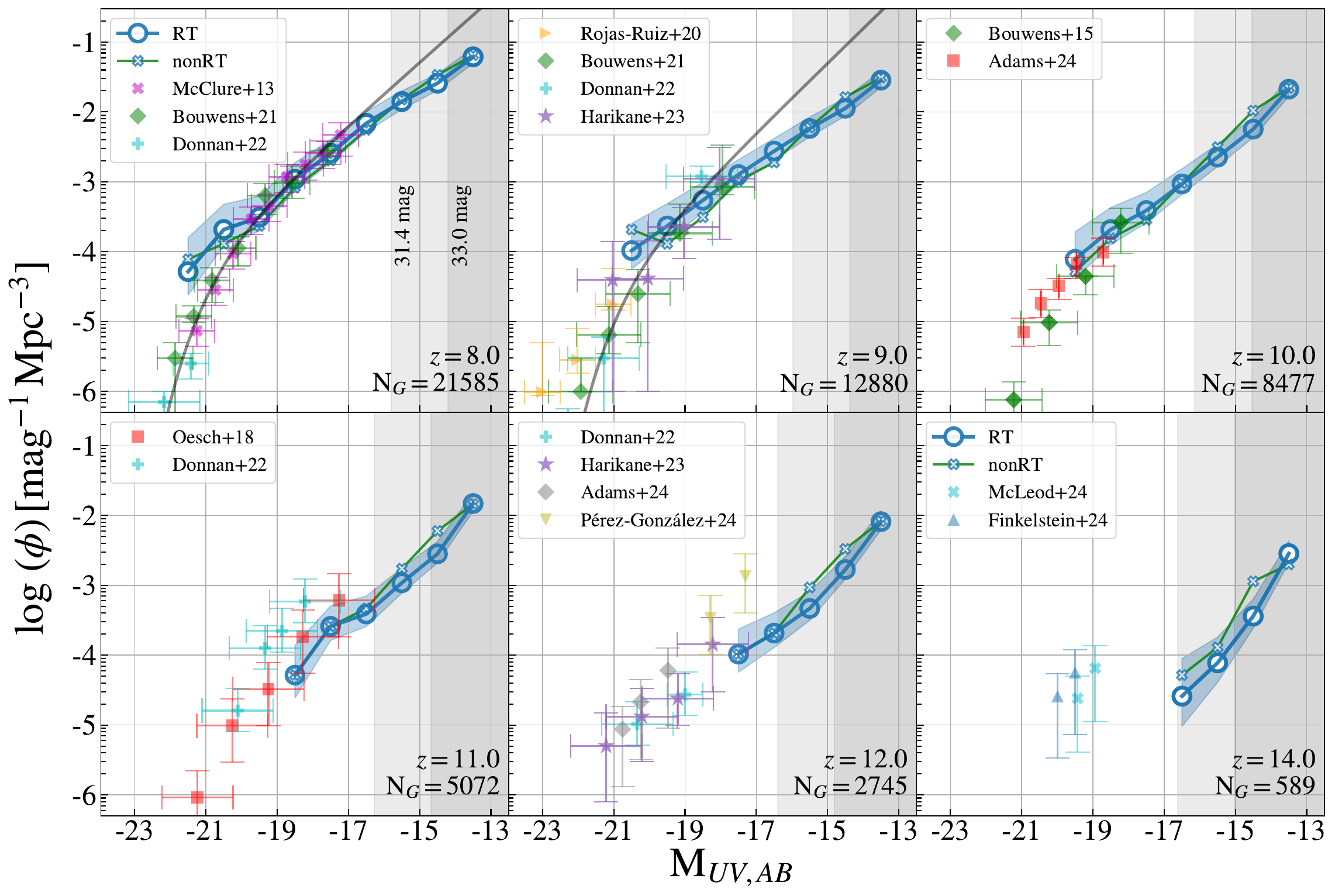}
\caption{The UV LFs derived from our simulations with 1$\sigma$ error bounds including both Poisson noise and sample variance. $N_G$ indicates the number of galaxies in the simulation at each redshift. Dark grey lines are Schechter fits from \cite{2016PASA...33...37F}. The LFs from the non-RT simulation are indicated with the green line and small, hollow crosses and essentially overlay the results from the RTsim. We also include observational LFs from \cite{2013ApJ...773...75O}, \cite{McLure2013}, \cite{2015ApJ...803...34B}, \cite{Oesch2018}, \cite{Rojas-Ruiz2020}, \cite{Bouwens2021}, \cite{Donnan2023} ($z\approx13$), \cite{Finkelstein24b}, \cite{McLeod24} ($z\approx13.5$), \cite{Perez23} ($z\approx12.25$), \cite{Adams2024}, and \cite{Harikane2023}.  The shaded areas indicate the regions where $m_{\rm UV} > 31.4$ mag (grey), a likely limiting magnitude for a JWST ultra-deep campaign and $m_{\rm UV} > 33$ mag (dark grey), a likely lensing limit. Our LFs are consistent with the observations at the bright end, and provide predictions at fainter luminosities.}
\label{fig:fit}
\end{center}
\end{figure*}

 Also, as discussed in $\S\ref{sec:fix}$, the earlier version of the subgrid model used in \cite{Sarmento2018} underestimated the resulting fraction of metal-free gas, as a function of time, resulting in stars with higher metallicities forming earlier. This was very likely a minor effect. Lastly, there was another processing-related error in \cite{Sarmento2018}: only a subset of the SPs were extracted and analyzed. Hence the MZR, LF and SFRD in that work are slightly underestimated and include an unresolved sampling bias.

We also include observationally derived points for $z=$ 8, 9, and 10 by \cite{2023ApJS..269...33N}.  The mass-metallicity relation for redshifts 8 and 9 agree well with these summary data points. We also agree with observations at z=10 but do not sample the function out to galaxies masses $M_\star > 10^{9}M_\odot$.

At galaxy masses $M_\star \le 10^7 M_\odot$ across $8 \le z \le 14$, $\big<Z_{\rm G}\big> \approx 8\times 10^{-2} Z_\odot$ for both simulations. The mean metallicity is larger when we consider $z < 11$ and $M_\star/M_\odot > 10^7$ where we see $\big<Z_{\rm G}\big> \gtrsim      0.1 Z_\odot$. The most massive galaxies with $M_\star/M_\odot \ge 10^8$ tend to an average of 0.2 to 0.3 $Z_\odot$. Conversely, at $z<11$ we see a larger range of metallicity for galaxies with $M_\star \le 10^6 M_\odot$. These small galaxies formed on the outskirts of enriched regions.

\subsection{Luminosity Functions}

Distant galaxies are characterized by their rest-frame UV fluxes, and galaxies with a high fraction of massive, hot, young stars are more luminous than those with an equivalent mass of lower-mass, older stars. Hence Pop III stars, in particular, have an outsized contribution to a galaxy's brightness. However, these hot stars burn out quickly and their impact on the galaxies' luminosities is short-lived. 

Figure \ref{fig:fit} depicts the simulations' UV LFs, at a rest-frame wavelength of $1500$\AA, along with observations across the redshift range depicted \citep{Bouwens2015,Bouwens2021,Donnan2023, 2016PASA...33...37F, Harikane2023, Finkelstein24b, McLeod24, Perez23, Adams2024, McLure2013, Oesch2018, Rojas-Ruiz2020,Adams2024}. Our  error estimates for the LFs include both the shot noise associated with the sample size and 1$\sigma$ errors due to cosmic variance \citep{Trenti2008}. Halo DM masses for the variance analysis were estimated via abundance matching. 

Fig.\ \ref{fig:fit} includes galaxies below $m_{\rm UV} = 33$ mag, a likely limit for JWST lensed fields \citep{Gardner_2006}. Our simulation volume of 38,624 cMpc$^{3}$ allows us to sample the LF, $\phi$,  to $\approx 10^{-4}\, \,{\rm mag}^{-1}\,{\rm cMpc}^{-3}$ on the bright end. Hence we typically do not sample the more rare but bright galaxies captured by the surveys depicted alongside our data. 
Our simulated LFs do not account for dust attenuation \citep[see][]{2017MNRAS.470.3006C, 2001PASP..113.1449C} due to the large uncertainties in dust modeling at these redshifts.

 Our LFs are in fair agreement with existing observations.  At $z=12$ the LF is approximately 0.5 to 1 magnitude dimmer than at $z=11$ at the equivalent galaxy densities. At $z=14$, an epoch that will likely push JWST observations into the lensing region, magnitudes fall off more quickly at the brighter end. Considering $z=12$ at $\phi=10^{-3}$,  the LF is a full magnitude dimmer at $z=14$ for the same density. At $z=12$ and $\phi=10^{-4}$ the same density of galaxies at $z=14$ are $\approx $2.1 magnitudes dimmer. 
Put another way, galaxy counts at the JWST detectability limits we posit, for both deep campaigns and lensed fields, are between 0.7 and 1.0 dex lower at $z=14$ as compared to $z=12$.  While rare, directly detectable (i.e. -- un-lensed) galaxies likely exist at these redshifts predicting their density would require a simulation at least 100 times the size presented here.

Comparing our RT and non-RT simulation we see almost no discernible differences. Radiative feedback is a minor effect, for our resolution, as compared to the IMF choice and metal mixing which determine the Pop III SFRD.  Note that this relative insensitivity to radiative transfer effects is likely limited to Pop III star formation in halos with masses above the atomic cooling limit, which is the focus of our study here. On the other hand, RT has a significant impact on minihalos with virial temperatures below $10^4,$K which are highly susceptible to photoevaporation \citep[\eg][]{Shapiro2004,Iliev2005}.  These halos have a strong impact on the progression of reionization itself, leading to additional recombinations that extend the transition from a neutral to an ionized intergalactic medium as shown in several recent studies \citep[e.g.][]{Daloisio2020,Ahn2021,Ocvirk2021,Chan2024}.  However, these effects occur below the limit at which they are likely to leave strong imprints in the post-reionization universe, as they are highly reduced for more massive galaxies which dominate the observed UV luminosity function and the cosmic star formation rate density \citep{WuX2019}. One possible complication, however, is the impact of high-redshift X-ray binaries, which can ionize atomic hydrogen and catalyzing H$_2$ formation \citep[e.g.][]{Fragos2013,Mirabel2011}.  This can lead to enhanced cooling, promoting additional early Pop III stars formation in mihalos \citep[e.g.][]{Jeon2014,Haemmerle2020}.

\begin{figure*}[t]
\begin{center}
\includegraphics[width=2\columnwidth]{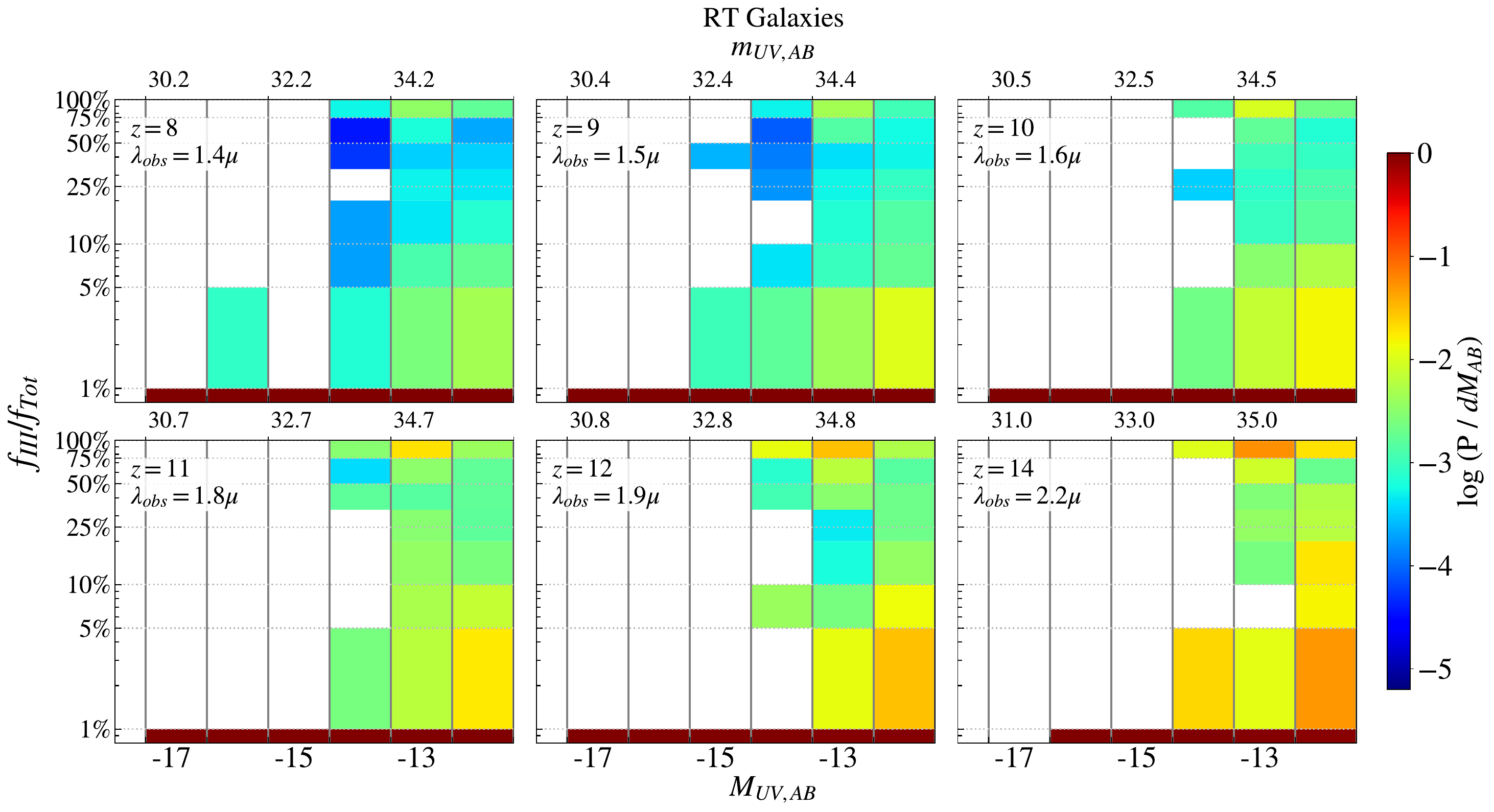}
\includegraphics[width=2.\columnwidth]{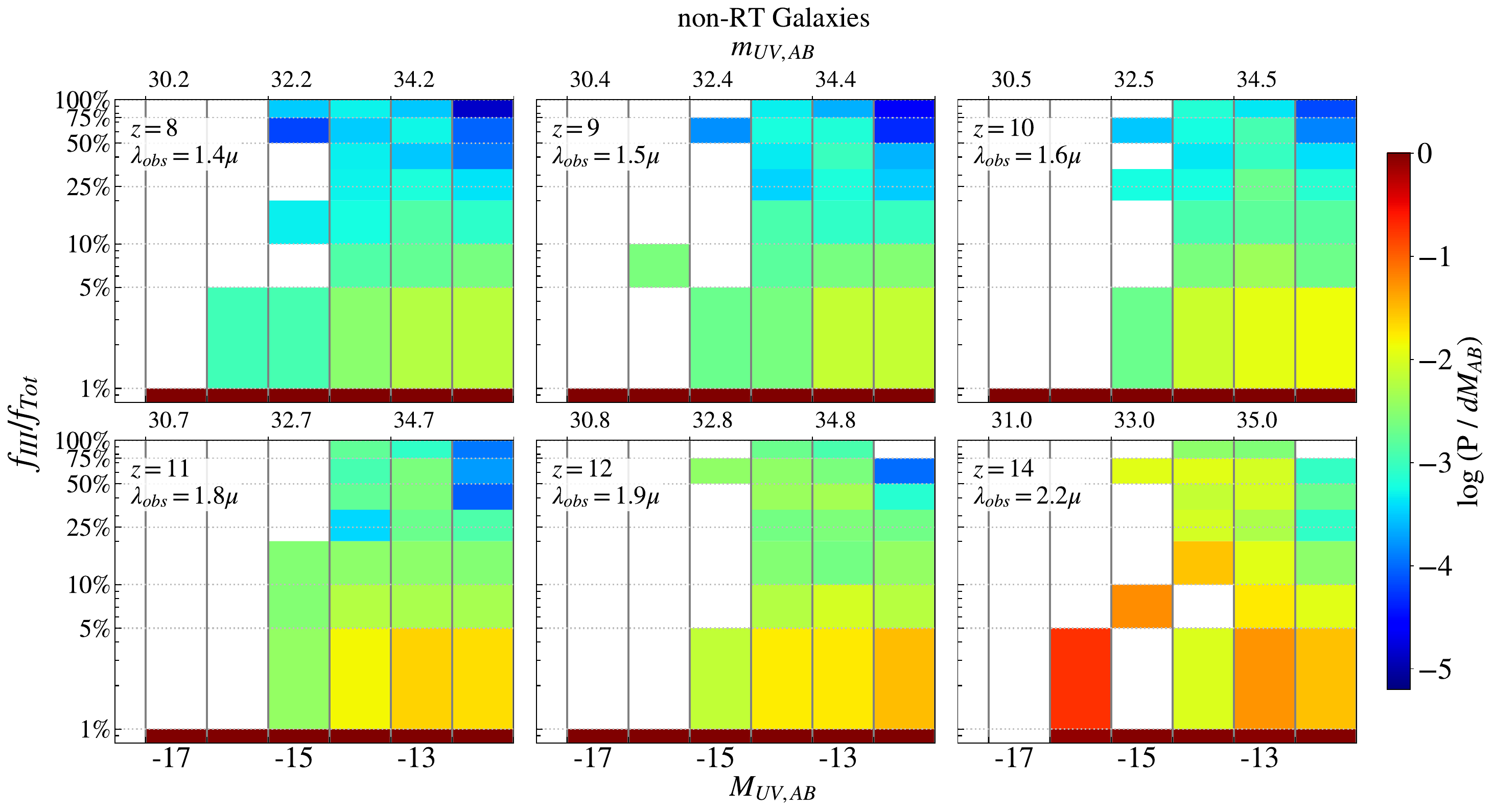}
\caption{The normalized probability, per magnitude bin, of finding a UV Pop III flux fraction, $\nicefrac{f_{\rm III}}{f_{\rm Tot}}$, for our galaxies in the RT simulation (\textit{Top})  and the non-RT simulation (\textit{Bottom} ).  Flux fractions less than 0.1\% are included in the bottom bin. Bin magnitudes are labeled at their right edge down to $M_{\rm UV} = -10$. The probability of finding RT galaxies with more than 0.1\% of their flux coming from Pop III stars is extremely low.  Even at $z=10$ and $M_{\rm AB} = -14$, a region accessible only in lensed fields, $< 0.1$\% of galaxies in our sample have $0.8\% \le f_{III}/f_{\rm Tot} \le 1.0\%$. The statistics improve little for the non-RT simulation. See the text for discussion.}
\label{fig:hist9}
\end{center}
\end{figure*}

\subsection{Pop III Flux}

Finally, we turn to the detectability of Pop III stars in high-redshift galaxy surveys. To quantify this, we compute the fraction of UV flux at $1500 \AA$ coming from Pop III stars for each of the galaxies in our simulation. While UV flux alone does not distinguish Pop III stars from Pop II stars, it allows us to determine the fraction of galaxies that are expected to have a hard-UV, He-ionizing flux, a signature that should be detectable by follow-up spectroscopy.

Figure~\ref{fig:hist9} depicts the probability of finding a given UV Pop III-generated flux fraction, normalized such that the probabilities sum to 1 {\it in each magnitude bin}, for both the RT and non-RT simulations. The lowest row of bins in these figures accumulates all of the galaxies with 1\% or less of their flux coming from Pop III stars. Magnitude bins are labeled at center while the bins on the y-axis are labeled at the top edge. 

 Here we see that RT galaxies with $M_{\rm UV, AB} \leq -15$ only appear at $z<10$. At $z=9$ there are only 2 galaxies with more than 1\% Pop III flux fraction (P3FF). One has a P3FF between 1 and 5\%, the other between 33 and 50\%. Considering the total number of galaxies in the simulation at this redshift, they each represent 1 galaxy in the sample of 12,880.  When considering only galaxies in the $M_{\rm UV, AB} = -15$ bin, they occur at a rate of 1 in 224 or less than 0.5\%: i.e. - more than 99\% of galaxies at $M_{\rm UV, AB} = -15$ at this redshift have a P3FF of $< 1\%$. 

 At $z=8$ there is a lone galaxy, one of 21,585 at this redshift, with $M_{\rm UV, AB} = -16$ ($m_{\rm UV, AB} = 31.2$) with 1-5\% of its flux coming from Pop III stars. To find RT galaxies with a significant (i.e. - $\geq 75\%$) P3FF occurring with a frequency of at least 1 in 100 we have to look to $z>9$ and to magnitudes $M_{\rm UV, AB} \geq -14$. These absolute magnitudes are likely beyond the detectability for JWST lensed fields \citep{Gardner_2006}. 

 In the RT simulation, at $z=8$, the probability of finding a galaxy with a P3FF $> 75\%$ with $M_{\rm AB} = -14$ is approximately $10^{-3.2}$ or a rate of approximately 1 in 1600. Contrast this result to \cite{Sarmento2019} where we found galaxies in the fiducial run, at $M_{\rm AB} = -14$, with more than 75\% of their flux with a probability of approximately $10^{-2.7}$. This is three times higher. This is a result of the Pop III Star Formation Rate Density (SFRD) in that work, which is seven times higher. Indeed, the overall SFRD in \cite{Sarmento2019} is $\approx 3$ times higher than the SFRDs in this work. While our simulations do not rule out the detection of brighter galaxies with more than a few percent of flux coming from Pop III stars, such galaxies are exceedingly rare given our assumptions.

 Chief among the assumptions affecting these latest results, specifically the SFRD and P3FF, are our decisions regarding feedback. As noted earlier, our decision to increase SN and stellar feedback helped to constrain star formation such that it is in reasonable agreement with observations, but also results in a lower overall SFRD as compared to our earlier works.

The situation is only marginally better when examining the non-RT simulation. For the same magnitude cutoff, $M_{\rm UV, AB} \leq -15$, at $z=14$ we see 3 galaxies with P3FF above 1\%. There is a lone galaxy with a P3FF of 1-5\% at $M_{\rm UV, AB} = -14$, one with a P3FF between 5-10\%, and a third between 50-75\%: both with $M_{\rm UV, AB} = -15$. Given the total number of galaxies in the simulation at this redshift, this represents an overall rate of less than 1\% of galaxies in the sample with $M_{\rm UV, AB} \le -15$ and a P3FF $ f_{III}/f_{\rm Tot} >1\%$. The direct detection of Pop III flux appears extremely difficult in light of these results. 

\section{Discussion and Conclusions}

The history of Pop III stars remains among the most compelling open questions in astrophysics. While these metal-free stars played a pivotal role in shaping the early universe, their detection has remained elusive, and even the best method to search for them remains unknown. In this study, we used large-scale cosmological simulations to better constrain the observability of these stars within high-redshift galaxies, accounting for the impact of radiative transfer (RT), turbulent mixing, and the IMF of Pop III stars. 

We carried out two 24 $h^{-1}$ comoving Mpc on a side cosmological adaptive mesh refinement simulations. The RT simulation combines self-consistent radiative transfer (RT) with our subgrid turbulent mixing model needed to accurately trace Pop III star formation.  The other simulation was identical to the first but did not include radiative transfer. Both simulations generate a star formation history in agreement with current observations but only about 1/3 of our earlier work \citep{Sarmento2019}. Also, the Pop III star formation rate density (SFRD) presented here are also somewhat lower than other simulations of the early universe. While some of this is due to resolution effects, as pointed out previously SN feedback and our scaling of radiative feedback likely play a role here.

The galaxy mass-metallicity relation for both simulations display a relatively high mean metallicity across the redshift and masses depicted. However, we do see a range of galaxy metallicities at masses $M_\star \le 10^6 M_\odot$  extending down to the mean seen in the FIRE simulations \citep{2024ApJ...967L..41M}. Our choice of a top heavy IMF for Pop III stars results in rapid pollution of early galaxies and partially mitigates the simulations' inability to resolve mini-halos -- a source of metals that likely contribute to the MZR in early structures. Current observations cannot confirm or refute our choice of Pop III IMF and hence the effect of these stars on the metallicity of early galaxies with $M_\star \le 10^7 M_\odot$ at $z\ge8$.

While observational constraints on the $z\ge 8$ galaxy luminosity function are still somewhat uncertain \citep{2016PASA...33...37F, 2015MNRAS.450.3032M, 2015ApJ...803...34B, 2015ApJ...808..104O, Adams2024}, our simulations match data where they are available. Our synthetic observations indicate that the LF faint end counts drop off by a factor of at least 5 when going from $z=12$ to $z=14$ and $M_{\rm AB} < -15,$ the likely detection limit for lensed JWST campaigns. These results display a slight reduction in the LF at $z \ge 10 $ as compared to the P3SN simulation in \cite{Sarmento2019}, a simulation with 1/8 the volume of this work. 
 
Like the overall Pop III SFRD, the fraction of Pop III flux in galaxies is reduced when using a top-heavy log-normal IMF.  In both \cite{Sarmento2018, Sarmento2019} (specifically the P3SN run in that later work), we always found Pop III galaxies with $f_{III}/f_{\rm Tot} > 75\%$. However,  fewer than 1 in 30, or $\approx 3\%$, of $M_{\rm UV,AB}$ = -15 galaxies, at $z=8$, had more than 75\% of their flux coming from Pop III stars in that work. In contrast, in our RT simulation, the probability of finding a $z = 8$, galaxy with $f_{III}/f_{\rm Tot} > 75\%$ with $M_{\rm UV,AB} = -14$, a full magnitude dimmer, is  $\approx 0.06\%$. This magnitude is likely below even JWST’s ability to detect.  Similarly, at $z=10$ and $M_{\rm UV,AB} \le -14$ less than $1\%$ of their flux comes from Pop III stars.

As previously discussed, the log-normal IMF postulated for Pop III stars in our simulations plays a significant role in these results.  Our choice of Pop III IMF results in only 1\% of primordial Pop III stars, by mass, surviving beyond 10 Myr. This results in $10\times$ more SN energy, and SN-generated metals, injected into the gas by the first generation of stars as compared to a similar mass of Pop II stars  that typically are assumed to follow a Salpeter IMF. This extra energy efficiently distributes metals into and around low mass primordial galaxies. Additionally, for the RT simulation, radiative heating provides more time for the metals to mix thoroughly into the gas. By contrast, assuming a Salpeter IMF for Pop III stars would result in more than 9 times the mass in Pop III stars surviving for more than 10 Myr across the redshift range studied.

This picture qualitatively agrees with \cite{2015MNRAS.452.1152J} who followed the feedback-regulated assembly of a single $10^{8}\, M_\odot$ halo in a zoom simulation using a high resolution (300 ckpc)$^3$ box.  They found that Pop III star formation was subdominant by $z = 13$ in this environment and negligible by $z=10,$ but noted that radiative transfer played an important role in determining these results.

Our results  indicate that radiative transfer is a subdominant component of simulations of early galaxy evolution,  for the galaxies we can resolve in this study. On the other hand, the details of our results are affected by our parameter choices. In particular, we boosted SN feedback energy by a factor of 5 in order to better match the SFRD to observations, and we assumed a local photon escape fraction $f_{\star, esc} =2$, such that 2x the number of stellar photons generated by stars escaped the local cell and were available for heating. These choices could be revisited as a function of simulation resolution. We plan to continue to explore the sensitivity of these results to such parameters in future studies. 

 There is little difference in galaxy metallicity or the Pop III flux fraction between the RT and non-RT simulations. As discussed, the adoption of a top-heavy IMF has significant ramifications for the fraction of long-lived Pop III stars and hence the detectability of a hard UV flux associated with them. Our results indicate that most galaxies on the edge of detectability, $M_{\rm AB} \approx -16$, have less than 1\% of their flux coming from Pop III stars modeled with a log-normal IMF with a characteristic mass of $60 M_\odot$. However, these result are also dependent on both SN and stellar feedback that modulates the SFRD. Since the Pop III stellar fraction scales with the overall star formation rate density (SFRD), simulation feedback tuning is a driver of these results and hence of predicted high-redshift galaxy observability, including galaxies' Pop III flux fraction. Feedback in cosmological simulations is an area of on-going research.

 Our modeling suggests that direct detection of Pop III-bright galaxies is unlikely if they are well described by our IMF and feedback prescriptions. Nevertheless, with the large survey volumes accessible to JWST, hope remains that astronomers will still capture the first clear indicators of rare primordial galaxies dominated by Pop III stars.

\begin{acknowledgments}
We would like to thank  Jennifer Chan, Huanqing Chen, Seth Cohen, Juna Kollmeier, and Rafaella Schneider for useful discussions. This work was supported by NASA Grant 80NSSC22K1265. The simulations and much of the analysis for this work was carried out on the NASA Aitken supercomputer. We would also like to thank the NASA High-End Computing Capability support team.
\end{acknowledgments}
\software{\textsc{Ramses-RT} \citep{2002A&A...385..337T}, MUSIC \citep{2013ascl.soft11011H}, pynbody \citep{2013ascl.soft05002P}}

\clearpage
\pagebreak


\bibliographystyle{aasjournal}
\bibliography{BigBib}

\end{document}